\documentclass[11pt]{article}
\usepackage{amssymb,epsf,graphicx}
\usepackage{amsmath}
\usepackage{epsfig}
\usepackage[numbers,sort&compress]{natbib}
\usepackage{latexsym}
\usepackage{graphicx}

\allowdisplaybreaks[1]

\begin{document}

\vskip 2.0cm

\centerline{\Large \bf Quantum gravity effects on
compact star cores
}

\vspace*{8.0ex}

\centerline{\large Peng Wang\footnote{E-mail: \tt
pengw@uestc.edu.cn}, Haitang Yang\footnote{E-mail: {\tt
hyanga@uestc.edu.cn}} and Xiuming Zhang\footnote{E-mail: {\tt
zhangxm@uestc.edu.cn}}}

\vspace{2.5ex}

\centerline{\large \it Department of Applied Physics,}
\vspace{1.0ex}

\centerline{\large \it University of Electronic Science and
Technology of China,}

\vspace{1.0ex}

\centerline{\large \it Chengdu, 610054, People's Republic of China}

\vspace{3.0ex}

\vspace*{10.0ex}

\centerline{\bf Abstract}
\bigskip
Using the Tolman-Oppenheimer-Volkoff equation and the equation of
state of zero temperature ultra-relativistic Fermi gas based on
generalized uncertainty principle (GUP), the quantum gravitational
effects on the cores of compact stars are discussed. Our results
show that ${2m(r)}/ {r}$ varies with $r$. Quantum gravity plays an
important role in the region $ r\sim 10^3 r_0$, where $r_0\sim
\beta_0 l_p $, $l_p$ is the Planck length and $\beta_0$ is a
dimensionless parameter accounting for quantum gravity effects.
Furthermore, near the center of compact stars, we find that the
metric components are $g_{tt}\sim r^4$ and
$g_{rr}=[1-{r}^2/(6r_0^2)]^{-1}$. All these effects are different
from those obtained from classical gravity. These results can be
applied to neutron stars or denser ones like quark stars. The
observed masses of neutron stars ($\leq 2M_\odot$) indicate that
$\beta_0$ can not exceed $10^{37}$, not as good as the upper bound
$\beta_0<10^{34}$ from simple electroweak consideration. This
means that incorporating either quantum gravity effects or nuclear
interactions, one obtains almost the same mass limits of neutron
stars.


\vspace{3.0ex}



\baselineskip=16pt

\vspace*{10.0ex}



The configuration of a spherically symmetric static star, composed
of perfect fluids, is determined by the Tolman-Oppenheimer-Volkoff
(TOV) equation (in c.g.s.
units) \cite{Tolman1939PR,Oppenheimer1939PR}
\begin{equation}
\frac{dP}{dr}=-(\rho+P/c^2)\frac{Gm(r)+4\pi
Gr^3P/c^2}{r[r-2Gm(r)/c^2]},\label{TOV1-PR}
\end{equation}
with
\begin{equation}
\frac{dm(r)}{dr}=4\pi r^2 \rho (r),\label{TOV1-mR}
\end{equation}
where $c$ is the velocity of light. $G$ is the gravitational
constant. $P$ and $\rho$ are respectively the pressure and the
macroscopic energy density measured in proper coordinates. Supplied
with an equation of state and appropriate boundary conditions, eqn.
(\ref{TOV1-PR}) and eqn. (\ref{TOV1-mR}) determine
$P(r)$, $m(r)$ and $\rho(r)$. If the pressure and gravitational
potential is everywhere small, i.e., $P(r)\ll \rho
c^2,\;2Gm(r)/c^2r\ll 1$, the TOV equation reduces to the
fundamental equation of Newtonian astrophysics
\begin{equation}
\frac{dP}{dr}=-\rho(r)\frac{Gm(r)}{r^2}.\label{TOV1-PR-New}
\end{equation}
Most of the low density compact stars like white dwarfs are well
described by Newtonian gravity. For compact stars like neutron stars
and other exotic compact stars, general relativity plays an
important role \cite{Weinberg1972GR}. The ideal neutron star is the
simplest model in which nuclear interactions are ignored and the
pressure of cold degenerate neutrons contends against the
gravitational collapse \cite{Oppenheimer1939PR}. There are basically
two ways to improve the model of compact stars. The first one is to
discuss more realistic structures of neutron stars and other Fermi
stars in theoretical and observational perspectives
\cite{AkmalP1997PRC,AkmalPR1998PRC,Engvik1996APJ,
Glendenning1999PRC,Muller1996NPA,Muther1987PLB,Prakash1987PRD,
Lattimer2001AA,Douchin2001AA,Hartle1978PReport}. In these works,
various types of equation of state (EOS) are introduced to represent
strongly interacting components and nuclear interactions. Nuclear
interactions significantly lift the maximum mass of neutron stars
from the Oppenheimer limit $0.7\;M_{\odot}$ to $2\;M_{\odot}$. A
more detailed discussion and references therein refer to
\cite{Lattimer2001AA}. Another direction is to introduce $f(R)$
theory or quantum gravity effects into the models\cite{
Santos2012,Deliduman2012,ArapogluJCAP2011,Pani2011PRD,Wiseman2002PRD,GERMANNI2001PRD}.
This way is of interest when addressing high density and high
pressure cold Fermi stars. This is the purpose of this paper. As a
first step in this direction, we adopt the ideal model without
nuclear interactions and the TOV equation.


In the absence of a full theory of quantum gravity, effective
models are useful tools to gain some features from quantum theory
of gravity. One of the most important models is the generalized
uncertainty principle (GUP), derived from the modified fundamental
commutation relation
\cite{Maggiore1993PLB65,Maggiore1993PLB83,Garay1995IJMPA145,
Kempf1995PRD1108,Kempf1992LMP1,Kempf1994JMP969,Kempf1994JMP4483}
\begin{equation}
[x,p]=i\hbar(1+\beta p^2),\label{1dGUP}
\end{equation}
where
$\beta=\beta_0l_p^2/\hbar^2=\beta_0/c^2M_p^2$, $l_p^2=G\hbar/c^3$, $M_p^2=\hbar
c/G$. $\hbar=h/2\pi$ is the Planck constant and $\beta_0$ is a dimensionless
parameter. With this modified commutator, one can easily derive the
generalized uncertainty principle (GUP)
\begin{equation}
\Delta x\Delta p\geq \frac \hbar 2 [1+\beta (\Delta
p)^2],\label{2dGUP}
\end{equation}
which in turn gives the absolutely smallest uncertainty in
positions, i.e., the minimum measurable length
\begin{equation}
\Delta x\geq
\Delta_{\textrm{min}}=\hbar\sqrt{\beta}=\sqrt{\beta_0}l_p.\label{GUPshixian-2}
\end{equation}
Note that the model in (\ref{1dGUP})  considers only the minimal
uncertainty in position. In this case, the quantum mechanics
structure underlying the GUP has been studied in full detail
\cite{Kempf1995PRD1108}. The statistics of ideal gases based on
GUP has been discussed by many authors
\cite{RAMA2001PLB103,LNC2002PRD125028,NozariMehdipour2007Chaos1637,
Fityo2008PLA5872,WYZ2010JHEP043}. In our recent work, we have
studied  a system composed of zero temperature ultra-relativistic
Fermi gas based on GUP \cite{WYZ2010JHEP043}. The Newtonian
equation with uniform pressure was employed to discuss stellar
structures. The proper particle number, energy density and
pressure  for an ultra-relativistic system were given in
\cite{WYZ2010JHEP043}
\begin{eqnarray}
\frac NV
&=&\frac{8\pi}{(hc)^3}E_H^3f(\kappa),\label{particlenumb1density}\\
\rho&=&\frac{8\pi}{c^2(hc)^3}E_H^4h(\kappa),\label{properENERGYDENSITY}\\
P&=&\frac{8\pi}{(hc)^3}E_H^4g(\kappa),\label{properpressure1}
\end{eqnarray}
where $ E_H=c/{\sqrt \beta}=M_pc^2/\sqrt{\beta_0}$ denotes the Hagedorn energy,
introduced in \cite{WYZ2010JHEP043} and
$\kappa=\varepsilon_F\sqrt{\frac{\beta}{c^2}}=\varepsilon_F/E_H$. Moreover
\begin{eqnarray}
h(\kappa)&\equiv&\frac 14\frac{\kappa^4}{(1+\kappa^2)^2},\label{hkapp1}\\
f(\kappa)&\equiv&\frac{1}{8}\left[\frac{\kappa(\kappa^2-1)}{(1+\kappa^2)^2}
+\textrm{tan}^{-1}(\kappa)\right],\\
g(\kappa)&\equiv&\kappa f(\kappa)-h(\kappa).\label{gkapadingy}
\end{eqnarray}
It is worth noting that when $\kappa$ increases, the proper
pressure blows up, while the proper energy density and the proper
number density are both bounded. This is a manifestation of the
minimal length.

The size of $\beta_0$ signals when quantum gravity effects enter
the story. In \cite{DasVagenas2008PRL221301}, based on the
precision measurement of Lamb shift, an upper bound of $\beta_0$
is given by $\beta_0< 10^{36}$. A relatively rough but stronger
restriction is estimated in \cite{BrauB2006PRD}. However, a better
bound is gained from simple electroweak consideration $\beta_0<
10^{34}$. For $\beta_0=10^{34}$, we rewrite eqns.
(\ref{properENERGYDENSITY}) and (\ref{properpressure1}) as
\begin{equation}
\rho=5.24\times 10^{95}\frac{1}{\beta_0^2}\;h(\kappa)\sim
10^{27}h(\kappa)\;\;(\textrm{kg}\cdot
\textrm{m}^{-3}),\label{densitykapa1}
\end{equation}
\begin{equation}
P=4.73\times 10^{112}\frac{1}{\beta_0^2}\;g(\kappa)\sim
10^{44}g(\kappa)\;\;(\textrm{Pascals}).
\end{equation}
Comparing these with the normal nuclear density $\rho_n=2.7\times
10^{17}\;\textrm{kg}\cdot \textrm{m}^{-3}$ and the pressure $P_n\sim
10^{34}\;\textrm{Pascals}$, the highest pressure recorded under
laboratory controlled conditions, we can find that in the vicinity
of nuclear matter equilibrium density, quantum gravitational effects
are not important. However, for density higher than the normal
nuclear one, it is of interest to investigate the cores of compact
stars like neutron stars and other exotic compact stars where
quantum gravity may play a leading role. As first approximation, we
consider only the degeneracy pressure regardless of the interaction
correction. On the other hand, to date, several accurate masses
determinations of neutron stars are available from radio binary
pulsars, as we will find that this may be used to constrain the
magnitude of $\beta_0$.

Two configurations of compact stars have been addressed in \cite{WYZ2010JHEP043},
by applying the Newtonian limit eqn. (\ref{TOV1-PR-New}) with uniform density.
One is that the star is almost composed of ultra-relativistic
particles. The other is that the major contribution to the mass is from
non-relativistic cold nuclei. However, to discuss the core of ultra-compact
stars like neutron
stars, one should use TOV equations
Setting $r=r_0\tilde{r},\,m=m_0\tilde{m}$, $P=P_0 \tilde{P}$ and %
\begin{equation}
\rho=\frac{m_0}{4\pi r_0^3}\tilde{\rho}\equiv \rho_0\tilde{\rho}
,\;P_0={\rho_0}{c^2},\;\frac{Gm_0}{c^2r_0}\equiv
1,\label{TOVscale-1}
\end{equation}
the TOV eqn.s (\ref{TOV1-PR}) and (\ref{TOV1-mR}) are reduced to the
following dimensionless ones
\begin{equation}
\frac{d\tilde{P}}{d
\tilde{r}}=-(\tilde{\rho}+\tilde{P})\frac{\tilde{m}+\tilde{r}^3\tilde{P}}
{\tilde{r}(\tilde{r}-2\tilde{m})},\label{tov-fermnoqg1dem}
\end{equation}
\begin{equation}
\frac{d\tilde{m}}{d\tilde{r}}=
\tilde{r}^2\tilde{\rho}.\label{tov-fermnoqg2dem}
\end{equation}

When there is no introduction of quantum gravity, for a system
almost composed of ultra-relativistic fermions, the
equation of state is  $\tilde{P}=\tilde{\rho}/3$. An exact solution
is given in \cite{Misner1964PRL635}
\begin{equation}
\frac{2\tilde{m}(\tilde{r})}{\tilde{r}}=\frac {3}{7},\;\;
\tilde{P}(\tilde{r})=\frac 1{14}\tilde{r}^{-2}.\label{noqgultracase}
\end{equation}
The pressure is not zero on the surface of the star. This does not
meet the physical boundary conditions. However, the
point is that it is  an analytic solution describing the central region of
compact stars with divergent pressure in the center \cite{Oppenheimer1939PR}.
Note that the length scale $r_0$ in eqn. (\ref{TOVscale-1}) is uncertain.
Thus $r$, $m$, $\rho$ and $P$
can be any size. %

From eqn. (\ref{noqgultracase}), the pressure is divergent in the center.
Therefore, influences from quantum gravity should be included in the discussion.
Obviously, near the surface, particles are non-relativistic
while in the region around the center, particles are
ultra-relativistic \cite{Oppenheimer1939PR}. This determines the equations
of state and boundary conditions. 

In the vicinity of $r=0$, the equation
of state is given by eqn. (\ref{particlenumb1density}), eqn.
(\ref{properENERGYDENSITY}) and eqn. (\ref{properpressure1}). Under the
limit $\kappa\to 0$, it is straightforward to recover $P = \rho/3c^2$.
Defining
$r=r_0\tilde{r},\,m=m_0\tilde{m}$ with
\begin{equation}
r_0^{-2}\equiv\frac{4\pi
G}{c^4}\frac{8\pi}{(hc)^3}{E_H^4},\label{r01minimall}
\end{equation}
\begin{equation}
m_0\equiv 4\pi r_0^3\frac{8\pi}{c^2(hc)^3}E_H^4=1.93\times
10^{-8}\beta_0\;\;(\textrm{kg}),
\end{equation}
\begin{equation}
P_0=\rho_0
c^2,\;\rho_0=\frac{8\pi}{c^2(hc)^3}{E_H^4},\label{TOVScale-2}
\end{equation}
where $r_0$ is the minimum radius in \cite{WYZ2010JHEP043}
\begin{equation}
r_0=\sqrt{\frac\pi 4}{\beta_0}l_p=\sqrt{\frac\pi 4}
\sqrt{\beta_0}\Delta_{\textrm{min}}=1.43\times
10^{-35}\beta_0\;\;(\textrm{m}).\label{r0scale1}
\end{equation}
Since $r_0$ in eqn. (\ref{r01minimall}) comes from eqn.
(\ref{TOV1-PR}), (\ref{TOV1-mR}), (\ref{properENERGYDENSITY}) and
(\ref{properpressure1}), $r_0$ represents the proper length. The
expressions (\ref{TOVScale-2}) and (\ref{r0scale1}) show  the system
can not be arbitrary scale, determined entirely by $\beta_0$. This
indicates
that our discussion is focused on the central region of compact stars. %
Substituting the above expressions for $P$ and $\rho$ (eqn.
(\ref{properENERGYDENSITY}) and eqn. (\ref{properpressure1})) into
eqn. (\ref{tov-fermnoqg1dem}) and eqn. (\ref{tov-fermnoqg2dem}), one
gets
\begin{equation}
\frac{d\tilde{m}(\tilde{r})}{d\tilde{r}}=\tilde{r}^2
h(\kappa),\label{TOV1-fermiddPR123}
\end{equation}
\begin{equation}
\frac{d\kappa(\tilde{r})}{d\tilde{r}}=\frac{-\kappa(\tilde{r})
\left[\tilde{m}(\tilde{r})+\tilde{r}^3g(\kappa)\right]}{\tilde{r}[\tilde{r}-2\tilde{m}(\tilde{r})]}.
\label{TOV1fereerddrer-PR}
\end{equation}


Since the density is regular in the center, one has $m(0)=0$ as a boundary condition.
After setting $\kappa_0\equiv\kappa(0)$ as another boundary condition,
eqn. (\ref{TOV1-fermiddPR123}) and eqn. (\ref{TOV1fereerddrer-PR})
are integrated numerically in Table I $-$ Table V.

In Table I to Table III, we perform the integration with different $\kappa(\tilde r)$.
Four conclusions can be drawn from these tables:
\begin{itemize}
  \item Different from the results obtained in classical gravity, ${2\tilde{m}(\tilde{r})}/{\tilde{r}}$
  varies with $\tilde r$ but not a constant $3/7$. For example, with $\kappa(\tilde r)=0.1$ in Table I, the deviation
  of ${2\tilde{m}(\tilde{r})}/{\tilde{r}}$ is about $4\%$.
  \item  ${2\tilde{m}(\tilde{r})}/{\tilde{r}}$ is not sensitive to different initial value $\kappa_0$.
  \item For large $\kappa(\tilde r)$ or small $\tilde r$, quantum gravity contribution
  is important to the value of ${2\tilde{m}(\tilde{r})}/{\tilde{r}}$. As $\kappa(\tilde r)$ decreases,
  or $\tilde r$ increases, the configuration approaches the classical
  one obtained in \cite{Misner1964PRL635}, with a constant ${2\tilde{m}(\tilde{r})}/{\tilde{r}}=3/7$.
  \item Quantum gravity plays an important role in the region $r\sim 10^3 r_0$.
\end{itemize}
Some analytic solutions can be obtained in extreme cases as follow.
\begin{itemize}
  \item Under $\kappa\to 0$,
  it is easy to see that $h(\kappa)\sim \kappa^4/4,\,g(\kappa)\sim \kappa^4/12$. Then
  from eqn. (\ref{TOV1-fermiddPR123}) and eqn. (\ref{TOV1fereerddrer-PR}), we
  obtain
  \begin{equation}
  \frac{2\tilde{m}(\tilde{r})}{\tilde{r}}=\frac {3}{7},\;\;
  \kappa(\tilde{r})=\left(\frac 67\right)^{1/4}\tilde{r}^{-1/2},\;\;
  \tilde{P}(\tilde{r})=\frac 1{12}\kappa^4=\frac 1{14}\tilde{r}^{-2},
  \;\;\;\;\;\;\textrm{for large}\;\; \tilde{r}.\label{littersoluti1}
  \end{equation}
  This solution is nothing but the classical one without
  quantum gravity.
  \item Under $r\to 0$ and $\kappa\to\infty$, eqn.
  (\ref{TOV1-fermiddPR123}) and eqn. (\ref{TOV1fereerddrer-PR}) can by
  replaced by asymptotic expressions
  \begin{equation}
  \frac{d\tilde{m}(\tilde{r})}{d\tilde{r}}=\frac
  1{4}\tilde{r}^2,\label{TOV1-fermiddPRasymptotic}
  \end{equation}
  \begin{equation}
  \frac{d\kappa(\tilde{r})}{d\tilde{r}}=\frac{-\kappa(\tilde{r})
  \left[\tilde{m}(\tilde{r})+\tilde{r}^3\frac\pi{16}\kappa(\tilde{r})\right]}
  {\tilde{r}[\tilde{r}-2\tilde{m}(\tilde{r})]}.
  \label{TOV1fereerddrer-PRasymptotic}
  \end{equation}
  The solution of these equations is
  \begin{equation}
  \tilde{m}(\tilde{r})=\frac{\tilde{r}^3}{12},\;\kappa(\tilde{r})=\frac{32}\pi
  \frac{1}{\tilde{r}^{2}},\;P(\tilde{r})=
  \frac{2}{\tilde{r}^{2}},\;\textrm{for}\;\tilde{r}\rightarrow
  0.\label{asymptosloulittelr}
  \end{equation}
  \end{itemize}

\noindent The solution (\ref{asymptosloulittelr}) represents the situation where quantum
gravity dominates. This happens near the center of ultra-compact stars. One can see that
it is quite different from the solution of classical gravity.
Table IV is the numerical result integrated for large $\kappa (\tilde r)$,
well consistent with the asymptotic solution
(\ref{asymptosloulittelr}).

For a spherically symmetric static compact star, the metric
is given by \cite{Weinberg1972GR}
\begin{equation}
g_{rr}\equiv
A(r)=\left(1-\frac{2Gm(r)}{rc^2}\right)^{-1}=\left(1-\frac{2\tilde{m}(\tilde{r})}{\tilde{r}}\right)^{-1}.
\end{equation}
\begin{equation}
g_{tt}\equiv-B(r),\;\;\frac 1B\frac{dB}{dr}=\frac{2G}{c^2r^2}\left[m(r)+\frac{4\pi r^3
P}{c^2}\right]\left[1-\frac{2Gm}{c^2r}\right]^{-1}.
\end{equation}
Then for $r\rightarrow 0$, from (\ref{asymptosloulittelr}), we have
\begin{equation}
A(r)=\frac 1{1-\tilde{r}^2/6},\;\;B(r)\sim
\tilde{r}^4.\label{ABQFREsults-1}
\end{equation}
One may compare (\ref{ABQFREsults-1}) with the classical results
\begin{equation}
A(r)=\frac 74,\;\;B(r)\sim \tilde{r}^{1/2}.\label{ABnotQFREsults-1}
\end{equation}

In Table V, eqn. (\ref{TOV1-fermiddPR123}) and eqn.
(\ref{TOV1fereerddrer-PR}) are integrated with a large initial
$\kappa_0$. It is interesting that $2\tilde{m}/\tilde{r}$ reaches
a maximum  value $0.734$ in the vicinity of $r=3.00\;r_0$. Our
calculation shows that near the center, $2\tilde{m}/\tilde{r}=
\tilde r^2/6$ which indicates that $2\tilde{m}/\tilde{r}$
increases with $\tilde r$. On the other hand, as $\tilde r\to
\infty$, $2\tilde{m}/\tilde{r}\to 3/7$. Therefore, the maximum of
$2\tilde{m}/\tilde{r}$ at $r=3.00\;r_0$ is a turning point, where
quantum gravity effect starts to dwindle. From Table V, one also
finds that $g_{rr}$ has a small range of fluctuation. A minimum
$(1-0.279)^{-1}=1.39$ is achieved at $r\simeq12.5\,r_0$.  This
minimum is about one-third of the maximum $(1-0.734)^{-1}=3.76$ at
$r\simeq 3.00\,r_0$. We do not have good explanation for this
fluctuation. It may be caused by the effectiveness of our model.
Finally, $g_{rr}$ tends to the constant $7/4$ at large $\tilde{r}$
as expected. The profile of $2\tilde{m}/\tilde{r}$ versus
$\tilde{r}$ is plotted in Fig \ref{FiG1}. One can see that the
upper limit, $8/9$ on the surface of a spherically symmetric
static star, is well satisfied.

Table VI shows the integrations from $\kappa_0=10 $ to the nuclear
density $\rho_n\simeq 10^{17}\;\textrm{kg}/\textrm{m}^3$ for
different $\beta_0$. The fifth line represents the values $\kappa$
corresponding to the nuclear density. The last two lines give the
masses (in solar mass units) and radii, when the stellar surface
density is taken as the nuclear density. From the second
conclusion drawn from Table I to Table III, the results are
insensitive to $\kappa_0$ provided $\kappa_0\geq 5$. Therefore,
the currently observed masses of neutron stars ($\leq 2M_\odot$)
indicates $\beta_0$ can not be greater than $10^{37}$. This
conclusion compatible with that from precision measurements of
Lamb shift. In another words, when incorporating the influence of
quantum gravity, one obtains the same mass limit of neutron stars
as that from considering nuclear interactions.

Table VII shows the integrations from $\kappa_0=10$ to
$\kappa=0.01$ for different $\beta_0$. In this region, quantum
gravity plays an important role. The last line lists $\rho
(\kappa=0.01)$ for different $\beta_0$, with reference to eqn.
(\ref{hkapp1}) and eqn. (\ref{densitykapa1}). In the region where
the density is less than $\rho(0.01)$, quantum gravity effects
almost have no effect. 
For $\beta_0=1$, the volume in which quantum gravitational effects
are important is in fact minuscule. Therefore, the observation of
quantum gravity effects depends only on the size of $\beta_0$. The
precise determination of the neutron star masses determines only
the upper limit of $\beta_0$.


In summery, we discussed the structure of ultra-compact star cores
by a simple effective quantum gravity model. The model, GUP,
introduces a new equation of state, determined by eqn.s
(\ref{particlenumb1density}), (\ref{properENERGYDENSITY}) and
(\ref{properpressure1}). By plugging the equation of state into
TOV equations, we found some different features from previous
works in literature.

Since quantum gravitational effects play an important role only in
high density, we considered configurations in which a star is
almost composed of ultra relativistic particles. The asymptotic
solutions near the center are given by (\ref{asymptosloulittelr})
and (\ref{ABQFREsults-1}). The complete picture is given by
numerical calculation. Quantum gravitational effects play a
leading role only in a relatively small range $\sim
10^3\;r_0=10^3\;\sqrt{\beta_0}\,\Delta_{\textrm{min}}$. Outside
this region, the solutions are determined by eqn
(\ref{noqgultracase}) and (\ref{ABnotQFREsults-1}). Our discussion
can be applied to neutron stars, for example. An upper bound of
$\beta_0 < 10^{37}$ was also achieved in Table VI. However, this
bound is larger than $\beta_0 < 10^{36}$, obtained from the
precision measurement of Lamb shift. On the other hand, simple
electroweak estimation gives a better bound $\beta_0 < 10^{34}$
than both of them. There are two ways to model compact stars. One
is including the nuclear interactions and another is to
incorporate quantum gravity effects.  It is of interest that our
results show that the two ways give the same mass limit of neutron
stars. It would be of importance in the future work to combine
both methods together in modelling compact stars. We hope the
refined models can further narrow the range of $\beta_0$.


\begin{table}
\begin{center}
\begin{tabular}{|r|c|c|c|c|}
\hline
$\kappa_0$ &$\kappa(\tilde{r})$&$m(\tilde{r})$ & $\tilde{r}$ &${2\tilde{m}(\tilde{r})}/{\tilde{r}}$  \\\hline%
1000.0 &   &  21.88  & 97.70& 0.448  \\  \cline{1-1} \cline{3-5}
100.0 &   &  21.87  & 97.70& 0.448  \\  \cline{1-1}   \cline{3-5}%
50.0  &   &  21.85  & 97.70& 0.447  \\   \cline{1-1}   \cline{3-5}%
20.0  &  &  21.79  & 97.60& 0.446  \\   \cline{1-1}   \cline{3-5}%
10.0  & 0.1 &  21.71  & 97.60& 0.445  \\  \cline{1-1}   \cline{3-5}%
8.0   &  &  21.65  & 97.50& 0.444  \\  \cline{1-1}   \cline{3-5}%
5.0   &   &  21.53  & 97.50& 0.442  \\ \cline{1-1}   \cline{3-5}%
3.0   &   &  21.29  & 97.40& 0.437  \\ \cline{1-1}   \cline{3-5}%
1.0   &   &  19.72  & 95.20& 0.414  \\ \cline{1-1}   \cline{3-5}%
0.5   &   &  18.36  & 87.90& 0.418  \\   
\hline%
\end{tabular}
\end{center}
\caption{\small Integration from $\kappa_0$ to
$\kappa(\tilde{r})=0.1$. The value of
${2\tilde{m}(\tilde{r})}/{\tilde{r}}$ is insensitive to initial
condition $\kappa_0$. ${2\tilde{m}(\tilde{r})}/{\tilde{r}}$ has
relatively large deviation from 0.429 since at $\tilde r=97.7$
quantum gravity has evident effects.}
\end{table}

\begin{table}
\begin{center}
\begin{tabular}{|r|c|c|c|c|} \hline %
$\kappa_0$ &$\kappa(\tilde{r})$&$\tilde{m}(\tilde{r})$ & $\tilde{r}$ &${2\tilde{m}(\tilde{r})}/{\tilde{r}}$  \\\hline %
1000.0 &    &  1985.47  &9250.70 & 0.429 \\  \cline{1-1}   \cline{3-5}%
100.0 &    & 1985.48    &9250.80 & 0.429  \\  \cline{1-1}   \cline{3-5}%
50.0  &    &  1985.51   &9251.00 & 0.429 \\  \cline{1-1}   \cline{3-5} %
20.0  &    &  1985.61   &9251.60 & 0.429  \\  \cline{1-1}   \cline{3-5} %
10.0  &    &  1985.76   &9252.50 & 0.429  \\  \cline{1-1}   \cline{3-5} %
8.0   &  0.01 &   1985.86  &9253.00 & 0.429 \\  \cline{1-1}   \cline{3-5}   %
5.0   &    &   1986.14   &9254.40 &   0.429 \\  \cline{1-1}   \cline{3-5} %
3.0   &    &   1986.72  &9257.20 & 0.429  \\  \cline{1-1}   \cline{3-5}%
1.0   &    &  1988.16   &9269.70 & 0.429 \\  \cline{1-1}   \cline{3-5}%
0.5   &    &  1981.62   &9265.40 & 0.428 \\  \cline{1-1}   \cline{3-5}%
0.1 &     &  2016.18  &9405.60& 0.429 \\ \hline     
\end{tabular}
\end{center}
\caption{\small Integration from $\kappa_0$ to
$\kappa(\tilde{r})=0.01$. The different initial value $\kappa_0$
has almost no effect on ${2\tilde{m}(\tilde{r})}/{\tilde{r}}$.
$\tilde r$ is large enough to overwhelm quantum gravity
influences.}
\end{table}



\begin{table}
\begin{center}
\begin{tabular}{|r|c|c|c|c|}%
\hline%
$\kappa_0$ &$\kappa(\tilde{r})$&$\tilde{m}(\tilde{r})$ & $\tilde{r}$ &  ${2\tilde{m}(\tilde{r})}/{\tilde{r}}$  \\\hline%
1000.0 &   &  198372.71  &925778.50& 0.429 \\\cline{1-1}   \cline{3-5} %
100.0 &  &  198372.88  &925778.70& 0.429 \\ \cline{1-1}   \cline{3-5}%
50.0 &   &  198373.07 &925778.90& 0.429 \\ \cline{1-1}   \cline{3-5} %
20.0 &   &  198373.65  &925779.50& 0.429 \\ \cline{1-1}   \cline{3-5}%
10.0 & 0.001 &  198374.58  &925780.30& 0.429 \\ \cline{1-1}   \cline{3-5}%
5.0 &   &  198376.38  &925781.40& 0.429 \\ \cline{1-1}   \cline{3-5}%
1.0 &   &  198393.76  &925802.80& 0.429 \\\cline{1-1}   \cline{3-5} %
0.1 &   &  198516.29  &925868.20& 0.429 \\ \cline{1-1}   \cline{3-5}%
0.01 &   &  201594.58  &940268.10& 0.429 \\\cline{1-1}   \cline{3-5} \hline%
\end{tabular}
\end{center}
\caption{\small Integration from $\kappa_0$ to
$\kappa(\tilde{r})=0.001$. 
The numerical results completely match the asymptotic solution
eqn. (\ref{littersoluti1}).}

\end{table}


\begin{table}
\begin{center}
\begin{tabular}{|r|c|c|c|c|}
\hline $\kappa_0$ &$\kappa(\tilde{r})$&$\tilde{m}(\tilde{r})$ & $\tilde{r}$&${2\tilde{m}(\tilde{r})}/{\tilde{r}}$  \\\hline 
1000.0 &   &  $2.85\times 10^{-2}$  & 0.700& $8.15\times 10^{-2}$   \\  \cline{1-1}   \cline{3-5} 
500.0  &    &   $2.77\times 10^{-2}$  & 0.693& $7.99\times 10^{-2}$  \\  \cline{1-1}   \cline{3-5} 
200.0  &  20 &  $2.52\times 10^{-2}$  & 0.672& $7.51\times 10^{-2}$   \\   \cline{1-1}  \cline{3-5} 
100.0  &   &   $2.14\times 10^{-2}$  & 0.636& $6.72\times 10^{-2}$  \\   \cline{1-1}  \cline{3-5} 
50.0   &    & $1.41\times 10^{-2}$  & 0.554& $5.10\times 10^{-2}$    \\  \cline{1-1}   \cline{3-5} 
30.0   &    &  $0.60\times 10^{-2}$  & 0.417& $2.89\times 10^{-2}$   \\  \cline{1-1 }  \cline{3-5} %
\hline
\end{tabular}
\end{center}
\caption{\small Integration from $\kappa_0$ to
$\kappa(\tilde{r})=20$. The large value of $\kappa$ corresponds to
$r\rightarrow 0$. ${2\tilde{m}(\tilde{r})}/{\tilde{r}}$ depends
sensitively on $\tilde{r}$.}

\end{table}


\begin{table}
\begin{center}
\begin{tabular}{|r|c|c|c|c|}
\hline $\kappa_0$ &$\kappa(\tilde{r})$&$\tilde{m}(\tilde{r})$ & $\tilde{r}$&${2\tilde{m}(\tilde{r})}/{\tilde{r}}$  \\\hline 
    & 100.0  &  $2.32\times 10^{-3}$  & 0.303& $1.53\times 10^{-2}$  \\   \cline{2-5} 
   & 50.0   &  $7.05\times 10^{-3}$  & 0.439& $3.21\times 10^{-2}$  \\   \cline{2-5} 
   & 20.0   &  $2.85\times 10^{-2}$  & 0.700& $8.14\times 10^{-2}$  \\   \cline{2-5} 
   & 10.0   &  $7.87\times 10^{-2}$  & 0.984& 0.160    \\\cline{2-5} 
   & 5.0    &  0.205                 & 1.365& 0.301  \\ \cline{2-5}
1000.0   & 1.0    &  0.983                 & 2.706& 0.727   \\ \cline{2-5} 
    & 0.8    &  1.100                 & 3.001& \textbf{0.734}  \\ \cline{2-5} 
    & 0.7    &  1.169                 & 3.220& 0.726  \\ \cline{2-5} 
   & 0.5    &  1.344                 & 4.052& 0.663 \\ \cline{2-5} 
  & 0.3    &  1.825                 & 8.091& 0.451  \\ \cline{2-5} 
   & 0.25    &  2.359                & 12.465& \textbf{0.279}  \\ \cline{2-5} 
    & 0.2    &  4.062                 & 22.801& 0.356 \\ \cline{2-5} 
   & 0.1    &  21.880                & 97.689& 0.448  \\\cline{2-5} 
&0.01    &  1985.47  &9250.70& \textbf{0.429}\\ \hline %
\end{tabular}
\end{center}
\caption{\small Integration with a fixed $\kappa_0=1000$.
$2\tilde{m}/\tilde{r}$ reaches its maximum value $0.734$ in the
vicinity of $r=3.00\;r_0$. $2\tilde{m}/\tilde{r}$ has a small
range of fluctuation and achieves a minimum value $0.279$ at
$r\simeq12.5\,r_0$. Eventually, $2\tilde{m}/\tilde{r}$ tends to
the constant $7/4$ at large $\tilde{r}$.}
\end{table}

\begin{table}
\begin{center}
\begin{tabular}{|c|c|c|c|}\hline
$\beta_0$ &$10^{37}$ & $10^{35}$ & $10^{33}$  \\\hline 
$\rho=5.24\times 10^{95}\frac{1}{\beta_0^2}h(\kappa)$ & $5.24\times
10^{21}h(\kappa)$  &  $5.24\times 10^{25}h(\kappa)$  & $5.24\times
10^{29}h(\kappa)$ \\ \hline %
$m_0=1.93\times 10^{-8}\beta_0 $ & $1.93\times 10^{29}\;\textrm{kg}
$ & $1.93\times 10^{27}\;\textrm{kg}$ & $1.93\times 10^{25}\;\textrm{kg}$ \\
\hline
$r_0=1.43\times 10^{-35}\beta_0 $ & $1.43\times 10^{2}\;\textrm{m} $
& $1.43\times 10^{0}\;\textrm{m}$ & $1.43\times 10^{-2}\;\textrm{m}$
\\ \hline
$\rho_n\simeq 10^{17}\;\textrm{kg}/\textrm{m}^3$ &
 $0.1$ & $0.01$ & $0.001$\\ \hline %
$M/M_{\odot}$ &$ 2.11$ & $1.93$ &$1.93$ \\\hline%
$R$ &$ 1.40\times 10^4\;\textrm{m}$ & $1.32\times 10^4\;\textrm{m}$
&$
1.32\times 10^4\;\textrm{m }$ \\\hline%
\end{tabular}
\end{center}
\caption{\small Integration from $\kappa_0=10 $ to the nuclear
density $\rho_n\simeq 10^{17}\;\textrm{kg}/\textrm{m}^3$ for
different $\beta_0$. The fifth line shows the values $\kappa$
corresponding to the nuclear density. The last two lines give the
masses (in solar mass units) and radii, when the stellar surface
density is taken as the nuclear density. The precise mass
determinations of neutron stars that have masses not larger than
$2M_\odot$ indicates  $\beta_0$ can not be greater than
$10^{37}$.}
\end{table}


\begin{table}
\begin{center}
\begin{tabular}{|c|c|c|c|c|}\hline
$\beta_0$&$10^{37}$ & $10^{36}$ & $10^{35}$ &$10^{34}$
\\\hline 
$m_0
$ & $1.93\times 10^{29}\;\textrm{kg}$
& $1.93\times 10^{28}\;\textrm{kg}$
&$1.93\times 10^{27}\;\textrm{kg}$&$1.93\times 10^{26}\;\textrm{kg}$
\\ \hline 
$r_0
$ & $1.43\times 10^{2}\;\textrm{m}
$
&$1.43\times 10^{1}\;\textrm{m} $
&$1.43\times 10^{0}\;\textrm{m} $&$1.43\times 10^{-1}\;\textrm{m} $ 
\\ \hline 
$M=1985.76\,m_0$ & $3.83\times 10^{32}\,\textrm{kg}$ & $3.83\times
10^{31}\,\textrm{kg}$
&$3.83\times 10^{30}\,\textrm{kg}$&$3.83\times 10^{29}\,\textrm{kg}$
\\ \hline 
$R=9252.50\,r_0$ &  $1.32\times 10^6\;\textrm{m}$ & $1.32\times
10^5\;\textrm{m}$
& $1.32\times 10^4\;\textrm{m}$& $1.32\times 10^3\;\textrm{m}$
\\ \hline 
$\rho(0.01)$ & $1.31\times 10^{12}\,\textrm{kg}/\textrm{m}^3$ &
$1.31\times 10^{14}\,\textrm{kg}/\textrm{m}^3$
&$1.31\times 10^{16}\,\textrm{kg}/\textrm{m}^3$&$1.31\times 10^{18}\,\textrm{kg}/\textrm{m}^3$
\\ \hline 
\end{tabular}
\end{center}
\caption{\small Integration from $\kappa_0=10$ to $\kappa=0.01$ for
different $\beta_0$. In this region, quantum gravity plays an
important role. The last line lists $\rho(\kappa=0.01)$ for
different $\beta_0$, with reference to eqn. (\ref{hkapp1}) and eqn.
(\ref{densitykapa1}). In the region where the density is less than
$\rho(0.01)$, quantum gravity effects is negligible. }
\end{table}

{
\begin{figure}
\centerline{\hbox{\epsfig{figure=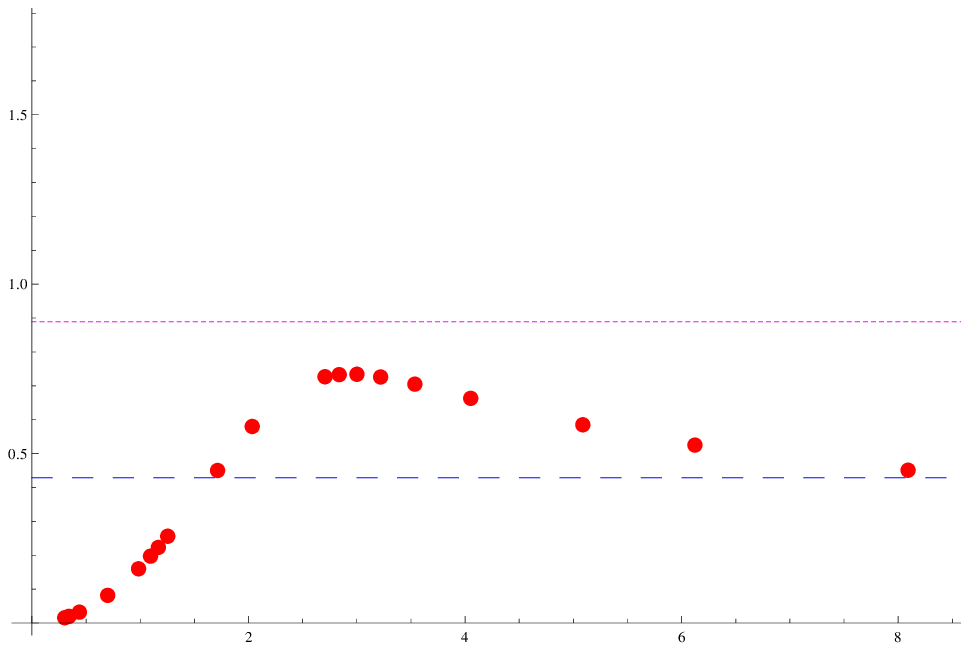, height=16cm}}}
\caption{For a fixed $\kappa_0=1000$, $2m/r$ versus the radius
$r=r_0\;\tilde{r}$. As $\tilde{r}\rightarrow 0$, $2m/r\sim r^2$.
$2m/r$  has a maximum around $\tilde{r}=3$. $2m/r$ acquires the
asymptotic value $0.429$ at large $r$. The dashed line represents
$2m/r=0.429$ while the dotted line represents $2m/r=8/9$, the
upper limit of $2m/r$ on the surface of a spherically symmetric
static star.} \label{FiG1}
\end{figure}}



\section*{Acknowledgement}

We are grateful to F. Lin for useful discussions and thank X. Guo
for help on numerical calculations. This work is supported in part
by NSFC (Grant No. 11175039 and 11005016).


\end{document}